\documentclass[11pt,fleqn,preprint]{article}
\usepackage{epstopdf} %converting to PDF
\usepackage{graphicx}
\usepackage[a4paper,left=2cm,right=2cm]{geometry}
\usepackage{amsmath}
\usepackage{float}
\usepackage[latin1]{inputenc}
\usepackage{amsmath}
\usepackage{amsfonts}
\usepackage{amssymb}
\usepackage{float}
\usepackage[toc,page]{appendix}
\floatstyle{boxed} 
\restylefloat{figure}

\newcommand{\be}{\begin{equation}}
\newcommand{\ee}{\end{equation}}
\newcommand{\bea}{\begin{eqnarray}}
\newcommand{\eea}{\end{eqnarray}}

\begin{document}
\title{Phenomenological consequences of introducing new fermions with exotic charges to $R(K^{(*)})$, muon (g-2), the primordial Lithium problem, and dark matter.\footnote{Preprint No. : HRI-RECAPP-2018-012}}
\author{Lobsang Dhargyal \\\\\ Previously @ Regional Center for Accelerator-based Particle Physics,\\\\\ Harish-Chandra Research Institute, HBNI, Jhusi, Allahabad - 211019, India}
%\date{July 31, 2018.}

%\preprint{HRI-RECAPP-2018-012}

\maketitle
\begin{abstract}

In this work we show that, by introducing two $SU(3)_{c}\times SU(2)_{L}$ singlet right handed fermions carrying opposite $U(1)_{Y}$ charges and while their left handed counterparts are singlet under $SU(3)_{c}\times SU(2)_{L}$ and neutral under $U(1)_{Y}$, in the regime where the new charged lepton masses are in the electroweak scale it will be able to explain the small neutrino masses via minimum-inverse seesaw scenario (MISS) as well as the reported $R(K^{(*)})$ and muon $(g-2)$ discrepancies. Also when the charged fermion masses are well above the electro weak scale the model can not explain the reported $R(K^{(*)})$ and muon $(g-2)$ discrepancies, but in this regime the model could explain the primordial Lithium problem. The model have another interesting side extension where it can produce a stable and singlet under strong interaction scalar baryon, provided the exotic fermions are vector like  under $U(1)_{Y}$ carrying fractional electromagnetic charges similar to $u_{R}$ and $d_{R}$ quarks (then MISS is not possible), which could constitute much of the dark matter mass of the universe which could link the origin of ordinary matter and DM. The model can also provide new annihilation channels for the scalar singlet DM as well as allowing a doubly charged scalar whose signatures could show up in HL-LHC, ILC, CEPC etc.

\end{abstract}
\maketitle

\section{Introduction.}

Standard-model (SM) of particle physics is the most successful theory of our understanding of the laws governing the natural world at subatomic to terrestrial scale where general relativity effects can be neglected. It is based on the symmetry group $SU(3)_{c}\times SU(2)_{L}\times U(1)_{Y}$ and its predictions has been verified and tested by many experiments over the last 40 or so years since its inception and no major deviation from its predictions has been found yet. However, there have been new developments since SM was proposed, one being the observations of neutrino oscillations (which in the simplest interpretation) indicating that neutrino have tiny but non-zero masses, which can be incorporated easily in SM by introducing three right handed neutrinos, but then SM can not give a satisfying answer to the reason why neutrino masses are much smaller than the masses of the other fermions in SM. Another of new developments since SM was the discovery of missing mass (DM) in astrophysical observations, to which SM has no candidate to account for. Then there are also recent reports of deviations from SM predictions in lepton universality observables in B decays as high as 4$\sigma$ in some cases besides the long standing disagreement between SM and experimental prediction in $(g-2)_{\mu}$ at the level of 3.6$\sigma$. Then there is also the so called primordial Lithium deficit problem in Big-Bang-Nucleus (BBN) synthesis where it is reported about $\frac{n_{Li^{+3}}}{n_{H^{+1}}} \approx \mathcal{O}(10^{-10})$ fewer $Li^{+3}$ nucleus observed then expected from BBN \cite{Li-rev}. Here in this work we will propose a new-physics (NP) model with new leptons and scalars and show that in the regime where masses of the new leptons are at the electroweak scale, the reported anomalies in $R_{K^{(*)}} = \frac{B \rightarrow K^{(*)} \mu^{+} \mu^{-}}{B \rightarrow K^{(*)} e^{+} e^{-}}$ \cite{simp-ref4}\cite{simp-ref2}\cite{simp-ref3}\cite{simp-ref5}\cite{simp-ref6}\cite{simp-ref7} and muon (g-2) \cite{PDG} can be explained and in the regime where the masses of the new leptons are well above the electroweak scale, the model can not explain $R_{K^{(*)}}$ and muon (g-2) but in this regime there are interesting possibilities of existence of neutral scalar baryons as well as scalar baryons carrying -3 electromagnetic charges, which can be candidate to account DM and primordial Lithium problem respectively. The paper is divide as follows, in Section:\ref{mod-det} we give the details of the model. In Section:\ref{Imp} we give the explanations of $R_{K^{(*)}}$ and muon (g-2) in the low mass regime of new leptons and in Section:\ref{MISS} we show that our model can explain the smallness of neutrino masses via minimal-inverse seesaw scenario (MISS). In Section:\ref{Li} we give the models implications to DM and primordial Lithium deficit in the high mass regime of the new leptons. And in Section:\ref{sect:conclusions} we conclude.

\section{Model details.}
\label{mod-det}

In standard-model(SM) we have left handed and right handed fermions carrying different non-zero $U(1)_{Y}$ charges, here we will introduce two new leptons whose right handed parts are charged under the $U(1)_{Y}$, and carries opposite $U(1)_{Y}$ charges to make the model free of axial gauge anomaly, while their left handed are required not to be charged under the $U(1)_{Y}$ plus we also introduce one lepto-quark and one inert-Higgs-doublet as shown in Table(\ref{tab1}). Then allowed Yukawa interactions are
\be
\mathcal{L}_{int} = \sum^{3}_{i=1}(h_{q_{i}}\bar{Q}_{iL}\phi_{LQ}f_{1R} + h_{i}\bar{L}_{iL}\eta f_{1R}) + \frac{m_{f}}{2}(\bar{f}_{1R}f_{2R}^{c} + \bar{f}_{2R}f_{1R}^{c}) + \sum_{i=1,j=1}^{3,2}y_{ij}^{\nu}\bar{L}_{iL}\hat{H}\nu_{jR} + \frac{1}{2}\bar{f}_{L}\mu f^{c}_{L} + \bar{f}_{L}M_{R}\nu_{R} + h.c
\label{Eqs.1}
\ee
where $f_{L} = (f_{1L}, f_{2L})^{T}$, $\nu_{R} = (\nu_{1R}, \nu_{2R})^{T}$ and it will be shown in section \ref{MISS} that the 2x2 matrices $M_{R}$ and $\mu$ along with 3x2 Dirac neutrino mass matrix from Higgs (H) term due to Higgs VEV can generate small neutrino masses via minimum-inverse seesaw scenario (MISS) \cite{miss}. In the basis $f_{R} = (f_{1R}, f_{2R}^{c})^{T}$ we can write $\frac{m_{f}}{2}(\bar{f}_{1R}f_{2R}^{c} + \bar{f}_{2R}f_{1R}^{c}) + h.c = m_{f}(\bar{f}_{1R}f_{2R}^{c}) + h.c$\footnote{where $(\bar{f}_{1R}f_{2R}^{c}) = (\bar{f}_{1R}C\bar{f_{2R}}^{T}) = -(f^{cT}_{1R}C^{T}C\bar{f_{2R}}^{T}) = -(f^{cT}_{1R}\bar{f_{2R}}^{T}) = (\bar{f}_{2R}f_{1R}^{c})$, the factor $\frac{1}{2}$ will be canceled, where $C$ is the fermionic charge conjugation operator which has the nice properties of $C^{-1} = C^{\dagger} = C^{T} = -C$ \cite{Zhi}.} as $\bar{f}_{R}M_{f}f_{R}$ where $M_{f} = m_{f}\begin{bmatrix} 0 & 1 \\ 1 & 0 \end{bmatrix}$.
\begin{table}[h!]
\begin{center}
\begin{tabular}[b]{|c|c|c|c|c|c|} \hline
Particles & L & $SU(3)_{c}$ & $SU(2)_{L}$ & $U(1)_{Y}$ & $\mathcal{Z}_{2}$ \\
\hline\hline
$f_{1 R}$ & 1 & 1 & 1 & -1  & -1 \\
\hline
$f_{1 L}$ & 1 & 1 & 1 & 0  & +1 \\
\hline
$f_{2 R}$ & -1 & 1 & 1 & +1  & -1 \\
\hline
$f_{2 L}$ & 1 & 1 & 1 & 0  & +1 \\
\hline
$\phi_{LQ}$ & 0 & 3 & 2 & 7/6 & -1 \\
\hline
$\eta$ & 0 & 1 & 2 & 1/2 & -1 \\
\hline
$\nu_{iR}$ & 1 & 1 & 1 & 0 & +1 \\
\hline
\end{tabular}
\end{center}
\caption{The charge assignments of new leptons under the SM gauge groups, lepton number (L) and $\mathcal{Z}_{2}$ with $i = 1, 2$.}
\label{tab1}
\end{table}
The mass $M_{f}$ can be diagonalized by $\frac{1}{\sqrt{2}}\begin{bmatrix} 1 & 1 \\ -1 & 1 \end{bmatrix}$ which gives two fermions of degenerate pseudo-Dirac masses $m_{f}$ with eigen functions $F_{1} = \frac{1}{\sqrt{2}}(f_{1R} + f_{2R}^{c})$ and $F_{2} = \frac{1}{\sqrt{2}}(-f_{1R} + f_{2R}^{c})$. We would like to point out here that if $f_{1}$ carry muon lepton number (which we assumed in this work unless explicitly stated otherwise) but $f_{2}$ need not carry muon lepton number but could carry electron lepton number (in this work) or tau lepton number, hence the phrase pseudo-Dirac masses. And $f_{1}$ carrying muon lepton number and $f_{2}$ carrying electron lepton number can explain why the anomalies are only observed in the muon sector instead of electron sector.

\section{Implications to $R_{K^{(*)}}$ and $(g-2)_{\mu}$.}
\label{Imp}

In general it is well known that flavor-changing-neutral-current (FCNC) observables are very sensitive to new-physics (NP) as SM is free of FCNC at tree level. Particularly the FCNC observables $R_{K^{(*)}} = \frac{Br(B \rightarrow K^{(*)}\mu^{+}\mu^{-})}{Br(B \rightarrow K^{(*)}e^{+}e^{-})}$ \cite{simp-ref4} are very well studied and so the LHCb \cite{simp-ref2}\cite{simp-ref3}\cite{simp-ref4}\cite{simp-ref5}\cite{simp-ref6} and Belle \cite{simp-ref7} reporting of about 4$\sigma$ deviation in observables related to these processes is very interesting \cite{simp-ref9}. A global fit to the data on various observables in these processes with a generic model independent Wilson coefficients gives the best fit Wilson coefficients as $C_{9}^{NP}$, $C_{9}^{NP} = -C_{10}^{NP}$ or $C_{9}^{NP} = -C_{9}^{'NP}$ with large $C_{9}^{NP}$ is preferred over SM $C_{9}^{SM}$ at the level of above 4$\sigma$ \cite{simp-ref8}\cite{simp-ref9}\cite{simp-ref10}. In our model, the NP will be able to generate the Wilson coefficients $C_{9}^{NP} = -C_{10}^{NP}$ via box loop diagrams, where a general model independent treatment of box loop contributions from new particles to $R_{K^{(*)}}$ is given in \cite{simp-ref11}. The range of NP Wilson coefficient at 1$\sigma$ in our type of models is given as \cite{simp-ref9}
\be
-0.81 \leq C_{9}^{NP} = -C_{10}^{NP} \leq -0.51\ (1\sigma).
\label{C9C10-Exp}
\ee
The key constrains on the NP parameters comes from the observables $Br(B_{s} \rightarrow \mu^{+}\mu^{-})$, $B^{0}_{s}-\bar{B}^{0}_{s}$ mixing and $b \rightarrow s \gamma$. The present bounds on these observables are \cite{simp-ref11}
\be
-2.1\times 10^{-5}\ TeV^{-2} \leq C_{B^{0}_{s}\bar{B}^{0}_{s}}(\mu_{H})^{NP} \leq 0.6\times 10^{-5}\ TeV^{-2}\ (2\sigma)
\label{CBB-exp}
\ee
and the constrain from $b \rightarrow s \gamma$ on $C_{7}^{NP}$ and $C_{8}^{NP}$ is given as \cite{simp-ref11}
\be
-0.098 \leq C_{7}^{NP}(\mu_{H}) + 0.24C_{8}^{NP}(\mu_{H}) \leq 0.07\ (2\sigma)
\label{BsGa-exp}
\ee
where $\mu_{H} = 2m_{W}$ and also with present measurements we have \cite{bll-exp}
\be
Br(B_{s} \rightarrow \mu^{+}\mu^{-})^{Exp.} = 2.8^{+7}_{-6}\times 10^{-9}
\label{bll-exp}
\ee
which is consistent with SM prediction so NP contribution to this observable is required to be small. There is also the reported anomaly in the muon (g-2) which is reported to be as high as 3.6$\sigma$ according to some estimate \cite{PDG} given as
\be
\delta a_{\mu} = a^{Exp}_{\mu} - a^{SM}_{\mu} = (288 \pm 63 \pm 49)\times 10^{-11}.
\ee
Now from Eqs.(\ref{Eqs.1}) we can have contributions from the inert-Higgs sector to the $\delta a_{\mu}^{NP}$ given as
\be
\frac{m_{\mu}^{2}|h_{\mu}|^{2}}{2\times 16\pi}\int^{1}_{0}dx[\frac{x^{2}-x^{3}}{m^{2}_{\mu}x^{2} + (m_{f}^{2}-m^{2}_{\mu})x + m_{H_{0}}^{2}(1-x)} + \frac{x^{2}-x^{3}}{m^{2}_{\mu}x^{2} + (m_{f}^{2}-m^{2}_{\mu})x + m_{A_{0}}^{2}(1-x)}]
\ee
where $H_{0}$ and $A_{0}$ are the CP even and CP odd neutral scalars respectively of the inert-doublet. For $h_{\mu} = 3$, $m_{f} = 160$ GeV, $m_{H_{0}} = 150$ GeV and $m_{A_{0}} = 300$, these values will be used as the benchmark values through out this work, we get $\delta a^{NP} = 2.055\times 10^{-9}$ which is within 1.4$\sigma$ of the reported deviation in this observable.\\
Now contributions to the $b \rightarrow s\mu^{+} \mu^{-}$ via box loop due to new particles in our model can be expressed as
\be
C_{9}^{NP} = -C_{10}^{NP} = N\frac{Re(h_{b}^{'}h_{s}^{'*})|h_{\mu}|^{2}}{2\times
32\pi \alpha_{EM}m_{f}^{2}}[S(x_{Q},x_{H_{0}}) + S(x_{Q},x_{A_{0}})]
\label{C9C10}
\ee
where $S(x,y)$ are Inami-Lin functions and given as \cite{Ina-Lin}\cite{simp-ref11}
\be
S(x,y) = \frac{1}{(1-x)(1-y)} + \frac{x^{2}\ln{x}}{(1-x)^{2}(x-y)} + \frac{y^{2}\ln{y}}{(1-y)^{2}(y-x)}
\ee
with $x_{Q} = \frac{m^{2}_{LQ}}{m^{2}_{f}}$, $x_{H_{0}} = \frac{m^{2}_{H_{0}}}{m^{2}_{f}}$ and $x_{A_{0}} = \frac{m^{2}_{A_{0}}}{m^{2}_{f}}$ where $m_{LQ}$ being mass of the Leptoquark involved and we take its benchmark value through out this work as $m_{LQ} = 900$ GeV which is above the present LHC lower bound and $\alpha_{EM} \approx \frac{1}{137}$ is taken here. The $h^{'}_{b}$ and $h_{s}^{'}$ are the quark sector Yukawa couplings in Eqs.(\ref{Eqs.1}) in the mass eigen state and we impose same conditions on the quark sector Yukawa couplings and CKM matrix elements as in \cite{ours3}\cite{ours4} where the angles of CKM matrix elements are fixed as $\pi \leq \theta_{12} \leq \frac{3\pi}{2}$ and $\frac{3\pi}{2} \leq \theta_{13}, \theta_{23} \leq 2\pi$, i.e the signs of the first two rows of CKM matrix elements are changed relative to the third row compared to the usual convention where all the angles are fixed in the first quadrant \cite{ours3}\cite{ours4}. Also similar to that in \cite{ours4} we impose
\be
h^{'}_{d} = -V_{ud}h_{d} - V_{cd}h_{s} + V_{td}h_{b} = 0
\ee
to satisfy the very stringent bounds from $K^{0}-\bar{K}^{0}$ and $B^{0}-\bar{B}^{0}$ oscillations which can be satisfied along with explaining the $R_{K^{(*)}}$ data for $h_{1s} = h_{1b} = \frac{2\sqrt{\pi}}{21.588}$ with $Re(h_{1d}) = 0.039$ and $Im(h_{1d}) = -5.51\times 10^{-4}$ \cite{ours4}. With these values of the Yukawa couplings in the quark sector we get $h^{'}_{b}h^{'*}_{s} = -0.027 + \mathcal{O}(10^{-5})i$, and taking the benchmark values of the masses and $h_{\mu}$ given before gives from Eqs.(\ref{C9C10})
\be
C_{9}^{NP} = -C_{10}^{NP} = -0.67,
\ee
which is within the 1$\sigma$ experimental bound given in Eqs.(\ref{C9C10-Exp}).\footnote{We have adopted this sign convention of the CKM matrix elements to satisfy Eqs.(10) and Eqs.(11) with Eqs.(11) requiring that $Re(h^{'}_{b}h_{s}^{'*})$ is negative, so the sign convention we adopted turns out to be one of the simplest way to satisfy all of them.... for details see \cite{ours3}} At this value of $h^{'}_{b}h^{'*}_{s}$ the NP contribution to $B_{s}^{0}-\bar{B}_{s}^{0}$ oscillation can be expressed as \cite{ours2}
\be
C_{B\bar{B}}^{NP} = \frac{(h_{b}^{'}h_{s}^{'*})^{2}}{128\pi^{2}m_{f}^{2}}[S(x_{Q},x_{Q})]
\ee
which give with benchmark values of the parameters $Re(C_{B\bar{B}}^{NP}) = 7.1\times 10^{-7}$ $TeV^{-2}$ which is about an order of magnitude smaller than the present 2$\sigma$ bound on this observable given in Eqs.(\ref{CBB-exp}). There is also contribution to the CP violation in $B_{s}^{0}-\bar{B}_{s}^{0}$ oscillation from imaginary part of $(h^{'}_{b}h^{'*}_{s})^{2}$, but due to smallness of the imaginary part of $(h^{'}_{b}h^{'*}_{s})^{2}$, NP contribution to CP violation in $B_{s}^{0}-\bar{B}_{s}^{0}$ oscillation is negligible, see \cite{simp-ref11}. Also with given values of the NP parameters, we get $C_{7}^{NP} + 0.24C_{8}^{NP} = -2.6\times 10^{-3}$ which is about two order of magnitude smaller than the present 2$\sigma$ bound on these Wilson coefficients given in Eqs.(\ref{BsGa-exp}) coming from $b \rightarrow s \gamma$ data \cite{simp-ref11}. With $C_{10}^{eff.} = C_{10}^{SM} + C_{10}^{NP}$ and $Br(B_{s} \rightarrow \mu^{+}\mu^{-})^{NP}$ being proportional to $|C_{10}^{eff.}|^{2}$ \cite{ours2} and with $C_{10}^{SM} = -4.31$ and $C_{10}^{NP} = +0.67$ gives $Br(B_{s} \rightarrow \mu^{+}\mu^{-})^{NP} = 2.6\times 10^{-9}$ which is well within 1$\sigma$ of the measured experimental bound given in Eqs.(\ref{bll-exp}). Contributions due to NP to $Br(Z \rightarrow (\bar{q}q))$ in the quark sector turn out to be negligible at the benchmark values of the parameters taken in this work, and $Br(Z \rightarrow \mu^{+}\mu^{-})^{NP} \approx Br(Z \rightarrow \bar{\nu}\nu)^{NP} \approx 1.25\times 10^{-6}$ for $m_{H_{0}} \approx m_{H^{\pm}}$ which is one and two orders of magnitude smaller than the present respective experimental bounds of $Br(Z \rightarrow \mu^{+}\mu^{-})^{Exp.}_{error} = 6.6\times 10^{-5}$ and $Br(Z \rightarrow \bar{\nu}\nu)^{Exp.}_{error} = 5.5\times 10^{-4}$. The $m_{H_{0}} \approx m_{H^{\pm}}$ assumption also avoid constrains from Peskin-Tekuchi $\Delta T$ and $\Delta S$ parameters as in this limit of inert-Higgs-doublet model $\Delta T^{NP} \approx 0$ and $\Delta S^{NP} \approx 0$ which is well within the present experimental bounds of $\Delta T^{Exp.} < 0.27$ and $\Delta S^{Exp.} < 0.22$ \cite{PDG}\cite{Pes-Tek}. Where in the above calculations we have taken the values of quark masses, CKM parameters and experimental bounds from PDG \cite{PDG} and for $Br(Z \rightarrow \bar{l}l(\bar{q}q))^{NP}$ we have used Eqs.(12) of \cite{ours4} with $(T_{3} - Q\sin{\theta_{W}}^{2}) \rightarrow (\sin{\theta_{W}}\tan{\theta_{W}})$, see also Eqs.(2.39) of \cite{zll}.

\section{Minimal Inverse Seesaw Scenario (MISS).}
\label{MISS}

With addition of only two right handed neutrinos to the SM, after Higgs developed and non-zero VEV, the last three terms in Eqs.(\ref{Eqs.1}) can be written as
\be
-\mathcal{L}_{m} = \bar{\nu_{L}}M_{D}\nu_{R} + \bar{f}_{L}M_{R}\nu_{R} + \frac{1}{2}\bar{f}_{L}\mu f^{c}_{L} + h.c
\ee
with $\mu$ being a Majorana 2x2 mass matrix along with $M_{D}$ being a 3x2 Dirac mass matrix generated due to non-zero Higgs VEV and $M_{R}$ being a 2x2 mass matrix between $f_{L}$ and $\nu_{R}$. The $\mu$ term being of Lepton number violating it is expected to be small which lead to small masses to the light  neutrinos, the well known inverse seesaw mechanism \cite{miss-ref16}. But our mass matrices correspond to the minimal inverse seesaw scenario (MISS) proposed recently in \cite{miss}. In the basis $(\nu_{L}, \nu^{c}_{R}, f_{L})$ the mass terms given above can be written in terms of a 7x7 symmetric mass matrix as \cite{miss}
\be
M_{\nu} = \begin{bmatrix} 0 & M_{D} & 0 \\ M_{D}^{T} & 0 & M_{R}^{T} \\ 0 & M_{R} & \mu \end{bmatrix}
\ee
where by redefinition of singlet fields $f_{L}$ and a unitary transformation on $\nu_{R}$ fields we can take $\mu$ as diagonal and $M_{R}$ as Hermitian without loss of generality \cite{miss}. Then with the usual conditions in inverse seesaw mechanism of $M_{R} > M_{D} >> \mu$, at the leading order in $M_{D}M_{R}^{-1}$ we can express the light neutrino mass matrix in MISS as
\be
m_{\nu} = M_{D}M_{R}^{-1}\mu (M_{R}^{T})^{-1}M_{D}^{T} = T\mu T^{T},
\ee
where $T = M_{D}M_{R}^{-1}$ is a 3x2 matrix. Then at $\mu \approx 1$ keV, $M_{D}M_{R}^{-1} = 10^{-2}$ we have the light neutrino masses in the order of 0.1 eV. This light neutrino mass matrix has been shown to be diagonalized by a unitary transformation as
\be
U^{\dagger}m_{\nu}U^{*} = diag(m_{1}, m_{2}, m_{3})
\ee
where $m_{1, 2, 3}$ being the light neutrino masses with one of the light neutrino is predicted to be massless. The heavy sector consist of pair of pseudo-Dirac neutrinos $P_{i} = (f_{iL}, \nu_{iR})$ with a tiny mass splitting between the corresponding CP conjugate Majorana components of order $\mu$. The left handed SM neutrinos can be expressed in terms of the mass eigenstates as
\be
\nu_{iL} \approx V_{ij\nu}\nu_{jL} + TU_{ijR}P_{jL}
\ee
where $U_{R}^{\dagger}M_{R}U_{R} = diag(m_{P_{1}}, m_{P_{2}})$ and $V_{\nu} \approx (1 - \frac{1}{2}TT^{\dagger})U$ and so effects of heavy neutrinos and non-unitarity of $V_{\nu}$ can be expressed in terms of SM charged current interactions of neutrinos as \cite{miss}
\be
\mathcal{L}_{CC} = -\frac{g}{\sqrt{2}}\bar{l_{L}}\gamma_{\mu}(V_{\nu}\nu_{iL} + TU_{R}P_{jL})W^{\mu} + h.c.
\ee
We direct the readers to \cite{miss} for a very comprehensive analysis of MISS and its phenomenological consequences in LFV, neutrino oscillations, constrains on non-unitary parameters and sensitivity to searches at neutrino factories etc.

\section{Scalar quarks, primordial Li problem and DM.}
\label{Li}

In the previous sections we have shown the implications of a particular way to realize the MISS model for neutrino mass generation to $R_{K^{(*)}}$ and $(g-2)_{\mu}$ in the mass regime of the new charged righted lepton pairs around electroweak scale. In this section we will analyze the regime of mass of the new charged righted lepton pairs well above the electroweak scale. When the mass of the new charged lepton is near TeV and above, contributions to the $R_{K^{(*)}}$ and $(g-2)_{\mu}$ from new particles becomes negligible and so it can not explain the discrepancies between present experimental data and SM predictions. Then the existence of $\phi_{LQ}$ and $\eta$ has lost its empirical basis and they need not exist at all. Also in what follows we will assume that $f_{1R}$ and $f_{2R}$ are just charged fermions which do not carry either lepton or baryon number contrary to that in the previous sections where they carry lepton number\footnote{this is aimed at leptogenesis (where some of lepton numbers get converted into baryons) so that we can have a clear sense of what is generated from what.....}. Then the new charged particles can have more exotic interactions if we introduce more $\mathcal{Z}_{2}$ odd scalars such that we can have Yukawa terms such as
\be
\begin{split}
-\mathcal{L}_{Y} = y_{0}\bar{l}^{c}_{R}f_{2R}\phi^{0} + y_{++}\bar{l}^{c}_{R}f_{1R}\phi^{++} + y_{4/3}\bar{d}^{c}_{R}f_{1R}\phi^{+4/3} + y_{1/3}\bar{u}^{c}_{R}f_{1R}\phi^{+1/3}\\
+ y_{2/3}\bar{d}^{c}_{R}f_{2R}\phi^{-2/3} + y_{5/3}\bar{u}^{c}_{R}f_{2R}\phi^{-5/3} + h.c
\end{split}
\label{Eq:Li}
\ee
where in Table(\ref{tab2}) we have shown the charges carried by the different scalars under the various symmetry groups.\footnote{in general, these exotic scalars can exist independent of existance of their fermionic counter parts...} Note that these terms can be added to Eqs.(1) but they have not much interesting concequences in the regime where the exotic fermion masses are small, unlike the heavy mass regime of the exotic fermions as will be shown in the following paragraphs.
\begin{table}[h!]
\begin{center}
\begin{tabular}[b]{|c|c|c|c|c|c|c|c|} \hline
Particles & L & B & $SU(3)_{F}$ & $SU(3)_{c}$ & $SU(2)_{L}$ & $U(1)_{Y}$ & $\mathcal{Z}_{2}$ \\
\hline\hline
$\phi^{0}$ & 0 & 0 & 1 & 1 & 1 & 0  & -1 \\
\hline
$\phi^{++}$ & -2 or 0 & 0 & 1 & 1 & 1 & +2  & -1 \\
\hline
$\phi^{+4/3}$ & -1 or 0 & -1/3 & $\bar{d}$ \& $\bar{s}$ & $\bar{3}$ & 1 & +4/3  & -1 \\
\hline
$\phi^{+1/3}$ & -1 or 0 & -1/3 & $\bar{u}$ & $\bar{3}$ & 1 & +1/3  & -1 \\
\hline
$\phi^{-2/3}$ & +1 or 0 & -1/3 & $\bar{d}$ \& $\bar{s}$ & $\bar{3}$ & 1 & -2/3 & -1 \\
\hline
$\phi^{-5/3}$ & +1 or 0 & -1/3 & $\bar{u}$ & $\bar{3}$ & 1 & -5/3 & -1 \\
\hline
\end{tabular}
\end{center}
\caption{The charge assignments of new scalars under the SM gauge groups, lepton number (L), Baryon number (B) and $\mathcal{Z}_{2}$. The $SU(3)_{F}$ denote the SM flavor symmetry of u, d and s quarks.}
\label{tab2}
\end{table}
Since $f_{1R}$ is similar to the SM right handed leptons, it is known that the terms with $\phi^{4/3}$ and $\phi^{1/3}$ (for leptoquarks) can be written with SM right handed leptons and quarks if they are made to be even under the $\mathcal{Z}_{2}$ group but then from the present mass limits ($m_{\phi} > m_{t}$) on such charged scalars (leptoquarks for SM leptons only) with such interactions \cite{PDG}, these scalars will be unstable. Also here due to $f_{2R}$ having opposite charge to that of a SM right handed leptons, the terms with $\phi^{2/3}$ and $\phi^{5/3}$ scalars are not allowed with SM leptons and quarks alone. As will be shown, presence of $\phi^{\pm 4/3}$ and $\phi^{\pm 1/3}$ has very interesting consequences if we assume that $m_{f} > m_{\phi}$ and in the baryongensis or leptogenesis process we have a production of a small amount of excesses of $\phi^{-4/3}(d/s)$ and $\phi^{-1/3}(u)$\footnote{where here the symbols u, d and s inside the bracket denote the SM $SU(3)_{F}$ flavor carried by the respective scalar quarks but in general the scalar $SU(3)_{F}$ need not be the same flavor $SU(3)_{F}$ in the SM quark sector.} along with SM quarks $u^{2/3}$ and $d^{-1/3}$ such that $\frac{n_{\phi\phi\phi}}{n_{H^{+}}} \approx \mathcal{O}(10^{-10})$ i.e the exotic baryon density ($n_{\phi\phi\phi}$) is much smaller than the SM baryon density ($n_{H^{+}}$). In general it is expected that the maximum density of the exotic baryons will be of the type $B(qq\phi)$ and  $B(q\phi\phi)$ but to determine the multiplets of these baryons, we need to know what the statistical law governing an exchange between a fermionic quark and a scalar quark should be, only then we will be able to determine the multiplets of such baryon states. However, under general assumptions, they could be in the flavor singlet, octat and decuplet states. Depending on the charge carried by the scalar baryons, these bound states could carry different charges and so have very interesting implications, for instance if they carry charge +1 like the proton, but much heavier than proton, then a small amount will be able to account much of DM mass by forming exotic hydrogen atoms which will have spectroscopic signature very similar to that of a normal hydrogen except corrections due to nucleus mass. If they exist they should show up in the LHC data but how much prominent their signature at LHC will be depends on their masses. It is expected that they all cascade down to $B(uu\phi)$ and $B(u\phi\phi)$ eventually.\\
Then exotic baryon with three scalar quarks $B(\phi\phi\phi)$ is also allowed whose color is in singlet combination and so it's flavor wave functions should be in either singlet combination  or octat conbinations or ducuplet combinations such that over all wave function is symmetric when any two of it's identical bosonic constituents are exchanged. One such scalar baryon $B(\phi\phi\phi)$ allowed have electromagnetic charge -3 which is of interest as possible solution to the primordial Li problem \cite{Li-rev}. The interesting baryon with charge -3 is $B(\phi^{-1/3}(u)\phi^{-4/3}(d,s)\phi^{-4/3}(d,s))$ (where u, d and s symbols denote the respective flavor carried by the scalar quark) carrying zero spin in the ground state and could be part of an $SU(3)_{F}$ flavor singlet state or octat states or decuplet states, and also due to the scalar quarks being bosons the size of these scalar baryons should be much smaller than the size of the proton and neutron and so  it's participation in the BBN is expected to be minimal due to reduce cross section for collision. Even presence of a very small amount of such exotic baryon could be able to explain the Li deficit, probably the ratio of the exotic baryon density to that of the density of the hydrogen nucleus at the order of $\frac{n_{\phi\phi\phi}}{n_{H^{+}}} \approx \mathcal{O}(10^{-10})$ will be enough. In what follows we will give a tentative argument regarding how the scalar baryons carrying electromagnetic charge of -3 ($B(\phi^{-1/3}(u)\phi^{-4/3}(d)\phi^{-4/3}(s))$) could explain the primordial Li deficit problem, we will not go through a detail analysis as it is very involved calculations and also it is beyond the scope of the present work. Given a hydrogen like atom formed between two charged particles of masses $m_{1} >> m_{2}$ carrying charges in unit of $e$ of $z_{1}$ and $z_{2}$ respectively then the Bohr energy levels are given as
\be 
E_{n} = -\frac{m_{2}}{m_{e}}z^{2}_{1}z_{2}^{2}\frac{m_{e}e^{4}}{32\pi^{2}\epsilon^{2}_{0}\hbar^{2}}\frac{1}{n^{2}} = -\frac{m_{2}}{m_{e}}z^{2}_{1}z_{2}^{2}E_{1}\frac{1}{n^{2}}
\label{Eq:ion}
\ee
where $E_{1} = \frac{m_{e}e^{4}}{32\pi^{2}\epsilon^{2}_{0}\hbar^{2}} = 13.6$ eV is the ionization energy of hydrogen atom.\footnote{also here we would like to point out that in a bound state of a heavy particle carrying EM charge -1(+1) with a proton (anti-proton), a transition fron 3rd Bohr energy level to the second Bohr energy level will emit a photon with energy 3.47 keV which could explain the observed gamma ray excess at 3.5 keV \cite{3.5keV}.}\\
Now the heavy $B(\phi^{-1/3}(u)\phi^{-4/3}(d)\phi^{-4/3}(s))$ with -3 charged ion ($z_{1} = 3$) can form a hydrogen like atom with a proton ($z_{2} = 1$) or with a Helium nucleus ($z_{2} = 2$) or with Lithium nucleus ($z_{2} = 3$) due to primary collisions, then the ionization energy of the such heavy atoms of the scalar baryon with a proton is about 224.746 keV compared to ionization energy of forming a hydrogen like atom with $Li^{+3}$ nucleus of 14.159 MeV from Eqs.(\ref{Eq:ion}) indicates that as the universe cools due to expansion, eventually at the time when temperature of the universe have dropped below the level where most of the photons have not enough energy to ionize the hydrogen like atom formed between $Li^{+3}$ and the heavy scalar baryon, still substantial amount of those photons will have enough energy to ionize the hydrogen like ion formed between the heavy scalar baryon and $He^{+2}$ and $H^{+1}$ and hence it is expected that eventually most of the -3 charged heavy scalar baryons will form a hydrogen like atom with a $Li^{+3}$ and so in that sense $Li^{+3}$ is expected to be below the prediction from that of cosmology with only SM quarks. If in deed such heavy atoms are formed then they absorption and emission spectral line of first excitation can be searched for in the X-ray band around $E_{2}(Li)-E_{1}(Li) = 10.619$ MeV. This absorption and emission line can be looked for by X-ray detectors such as Chandra-X, HXMT, XRISM etc. although they are expected to be very faint compare to normal hydrogen absorption and emission line by a factor of about $\frac{n_{Li}}{n_{H}} = 10^{-10}$. The simplest collider signatures would be $pp/ee \rightarrow g^{*}/Z^{*}/\gamma^{*} \rightarrow \bar{\phi}(q)\phi(q) \rightarrow (\bar{q}q)^{*} \rightarrow hadronic\ final\ states$, which could show up in the form of a heavy resonance at invariant mass of $\bar{\phi}(q)\phi(q)$ such as e.g $pp/ee \rightarrow g^{*}/Z^{*}/\gamma^{*} \rightarrow \bar{\phi}(s)\phi(s) \rightarrow (\bar{s}s)^{*} \rightarrow K^{+}K^{-}$ or $\bar{K}^{0}K^{0}$, etc.\\
Another very interesting observation to be made about Eqs.(\ref{Eq:Li}) is that instead as done above where $Q_{f_{1R}} = -Q_{f_{2R}} = -e$ if we introduced two new vector like fermions carrying same Y as $d_{R}$ quark and $u_{R}$ quark as $Q_{f_{1}} = -1/3e$ and $Q_{f_{2}} = 2/3e$ respectively with both odd under $Z_{2}$ and singlets under $SU(3)_{c}\times SU(2)_{L}$ and $m_{f_{1,2}} > m_{\phi(d,u)} + m_{d,u}$.\footnote{of course now MISS mechanism is not possible with left handed fermions charged....} Then the corresponding allowed Yukawa terms  are $y_{1}\bar{d}_{R}f_{1L}\phi(d) + y_{2}\bar{u}_{R}f_{2L}\phi(u)$ and the respective scalar quarks $\phi(u)$, $\phi(d)$ and $\phi(s)$ are required to be neutral and stable which implies the scalar baryon $B(\phi(u)\phi(d)\phi(s))$ will be also neutral, stable as well as singlet under the strong interaction and could be a DM candidate.\footnote{also with $\phi(u)$ carrying $Q = +2/3$ and $\phi(d,s)$ carrying charges $Q = -1/3$ similar like the u and d type quarks then $B(\phi(u)\phi(d)\phi(s))$ will be neutral and singlet under strong interaction and could be DM although its stability under EM is not clear as the constituents being scalars unlike neutron whose constituents are fermions, but baryon number conservation could save it.} This scalar baryon is stable under strong color interaction due to asymptotic freedom but it may not be stable under its own gravity except baryon number conservation stabilizing it! However if the exotic fermions carry charges $Q_{f_{1}} = -1$ and $Q_{f_{2}} = 0$ then all three scalar quarks above carry $Q = +\frac{2}{3}$ and we can have a stable scalar baryon of charge +2, and if they are very heavy ($\mathcal{O}(TeV)$) then a small amount ($\frac{n_{\phi\phi\phi}}{n_{H}} \approx \mathcal{O}(6\times 10^{-3})$) could account much of DM mass of the universe, whose transition spectra will be very close to that of Helium atom except due to corrections from difference in nucleus masses and also similar to Helium atom this exotic Helium is chemically inert as well.\footnote{also $Q_{f_{1}} = +1/3$ and $Q_{f_{2}} = +4/3$ will generate scalar baryon of charge -2 could absorb about $\frac{n_{\phi\phi\phi}}{n_{H}} \approx \mathcal{O}(6\times 10^{-3})$ of protons or Helium and account for the DM mass.......} This baryon will be stable under it's own gravity due to electromagnetic repulsions among its constituents\footnote{These scalar baryons are very different from neutron or helium nuclei as this scalar baryon is expected to be much smaller in size than neutron or helium nuclei and so would not have participated much in the BBN.}.\footnote{in general if we take the size of these scalar baryons in scale of the proton or the neutron then since the scalar baryons are much heavier and so much denser they are expected to sink to the core of stars and galaxies and planets etc. and so they spectral signatures may not be that easy to find.....}\\
The other interesting thing about the Eqs.(\ref{Eq:Li}) is the first term in which if $m_{f} > m_{\phi^{0}}$ then the scalar $\phi^{0}$ is the scalar singlet DM candidate, for which the Higgs portal has been ruled out due to small Yukawa coupling imposed by the direct and indirect measurements except near the Higgs resonance neck region with Yukawa coupling of $\mathcal{O}(10^{-4})$, as such small Yukawa coupling predicts over abundance of DM relic density \cite{GAMBIT} in low mass region. However here the scalar singlet has the new Yukawa term which provide new annihilation channels and could avoid over abundance problem in the regime where the scalar singlet DM masses are below 100 GeV \cite{sadhukhan}.\\
Then the doubly charged scalar term is also interesting in a sense that if it is very heavy but $m_{f_{1R}} > m_{\phi^{\pm\pm}}$ then it is stable, and if some how a small excess of $\phi^{++}$ is generated over the $\phi^{--}$ in the early universe, then bound state of $\phi^{++}$ with two electron will be chemically inert and have absorption and emission spectra very close to the absorption and emission spectra of helium atom and so they may have been counted as helium but since they are much more massive then helium, they could actually account much of the universe's invisible mass (DM). For a recent collider signature analysis of doubly charged scalars with SM fermions see \cite{doubly-chrge}. For completeness we would like to point out that, instead of heavy stable scalars, if the heavy fermions are stable, then also most of the conclusion drawn above in section 5 are true, with major difference being the total spin which can be half integral here and have observable effects in the spectroscopic signatures related to spin effects. Also most of the statements regarding spectroscopic observables and their consequences shown in section 5 are true even if the particles are not composite but fundamental.\footnote{side note: another interesting observation is that, if there exist very weakly interaction and light exotic neutral stable vector bosons (singlet under SM local gauge groups), composite or other wise, such that there is a cosmic back ground of them similar to the CMB, if they are not U(1) gauge bosons but just vector bosons, then they can have vector like mix interaction with different quark and lepton flavor states and their temperature close to CMB temperature could explain neutrino oscillation and also they could contribute to the DM mass depending on their masses....... and also suppose nature only use the global gauge symmetry (in the sense of BRS symmetry : re-normalizability requirement) but not the local gauge symmetry (i.e after detail analysis in future, it turns out that the 2012 LHC scalar is not SM Higgs and SM Higgs does not exist at all), then there is the possibility of new exotic scalar interaction terms(also tensor interactions) mixing SM fermionic flavors and a cosmic background of them can induce neutrino oscillation and dark matter effects as well..... also  even with the SM gauge symmetry, a scalar (or even tensor) doublet with very light and stable neutral component can do the job too....}

\section{Conclusions.}
\label{sect:conclusions}

In this work we have presented an extension of SM by introducing two new leptons whose right handed parts are singlet under the SM non-Abelian gauge symmetries but carries opposite charges under the $U(1)_{Y}$ while their left handed is singlet under the whole SM gauge group. We have shown that, by introducing different kind of new scalars with respect to whether the masses of the new charged fermions are near the SM energy scale (low) or well above SM energy scale (high), our extension can explain the $R_{K^{(*)}}$ and muon (g-2) in the low mass regime and it could explain the primordial Lithium problem in the high mass regime beside its neutral left handed parts able to generate small neutrino masses via MISS. Also a very interesting side observation we made is that there are different ways of assigning charges if the new fermions are assumed to be vector like (instead of chiral as assumed in most part of this work) under $U(1)_{Y}$ then we can have stable scalar baryon that could constitute a large part of universe's invisible mass (DM).

%\newpage

{\large Acknowledgments: \large} This work was partially supported by funding available from the Department of Atomic Energy, Government of India, for the Regional Centre for Accelerator-based Particle Physics (RECAPP), Harish-Chandra Research Institute.


\begin{thebibliography}{99}

\bibitem{miss} M. Malinsky, T. Ohlsson, Z. Z. Xing and H. Zhang, \textsl{arXiv: 0905.2889v2 [hep-ph] 19 Aug 2009}

\bibitem{simp-ref4} R. Aaij et al. (LHCb), \textsl{Phys. Rev. Lett. 113 (2014) 151601}

\bibitem{simp-ref2} J. Matias, F. Mesicia, M. Ramon and J. Virto (LHCb), \textsl{JHEP 04, 104 (2012), 1202.4266}

\bibitem{simp-ref3} R. Aaij et al. (LHCb), \textsl{Phys. Rev. Lett. 111 (2013) 191801}

\bibitem{simp-ref5} S. Descotes-Genon, T. Hurth, J. Matias and J. Virto (LHCb), \textsl{JHEP 1305, 137 (2013), 1303.5794}

\bibitem{simp-ref6} R. Aaij et al. (LHCb), \textsl{JHEP 02, 104 (2016)}

\bibitem{simp-ref7} A. Abdesselam et al. (Belle), \textsl{LHCSki 2016 Abergurgl, Austria, April 10-15, 2016}

\bibitem{simp-ref9} S. Descotes-Genon,L. Hofer, J. Matias and J. Virto (2015), \textsl{arXiv: 1510.04239}

\bibitem{simp-ref8} W. Altmannshofer amd D. M. Straub, \textsl{arXiv: 1503.06199}

\bibitem{simp-ref10} T. Hurht, F. Mahmoudi and S. Neshatpour, \textsl{arXiv: 1603.00865}

\bibitem{Ina-Lin} T. Inami and C. S. Lim, \textsl{Prog. Theor. Phys. 65 (1981) 297, [Erratum: Prog. Theor. Phys. 65.1772 (1981)]}

\bibitem{simp-ref11} P. Arnan, Lars Hofer, F. Mescia and A. Crivellin (2017), \textsl{DOI: 10.1007/JHEP04(2017)043.}

\bibitem{PDG} M. Tanabashi et al. (Particle Data Group), \textsl{Phys. Rev. D 98, 030001 (2018)}

\bibitem{Pes-Tek} M. E. Peskin and T. Takeuchi, \textsl{Phys. Rev. Lett. 65, 964 (1990)}

\bibitem{bll-exp} V. Khachatryan et al. (LHCb, CMS), \textsl{Nature 522 (2015) 68-72}

\bibitem{zll} C. W. Chaing, H. Okada and E. Senaha, \textsl{Phys. Rev. D 96 (2017) no.1, 015002}

\bibitem{ours3} L. Dhargyal and S. K. Rai, \textsl{arXiv:1806.01178}

\bibitem{ours4} L. Dhargyal, \textsl{arXiv:1808.06499}

\bibitem{ours2} L. Dhargyal, \textsl{arXiv:1711.09772}

\bibitem{miss-ref16} R. N. Mohapatra and J. W. F. Valle, \textsl{Phys. Rev. D 34, 1642 (1986)}

\bibitem{Li-rev} B. D Fields, \textsl{arXiv: 1203.3551v1 [astro-ph.CO] 15 Mar 2012}

\bibitem{GAMBIT} The GAMBIT Collaboration, \textsl{DOI:10.1140/epjc/s10052-017-5112-1}

\bibitem{sadhukhan} D. Borah, S. Sadhukhan and S. Sahoo, \textsl{Physics Letter B 771 (2017) 624-632}

\bibitem{doubly-chrge} V. A. Gani, M. Y. Khlopov and D. N. Voskresensky, \textsl{arXiv: 1808.06816v1 [hep-ph] 21 Aug 2018}

\bibitem{Zhi} Zhi-zhong Xing, \textsl{https://doi.org/10.1016/j.nuclphysbps.2010.08.006}

\bibitem{etc1} Basabendu Barman, Debasish Borah, Lopamudra Mukherjee, Soumitra Nandi, \textsl{https://arxiv.org/abs/1808.06639}

\bibitem{etc2}  A. K. Alok, B. Bhattacharya, A. Datta, D. Kumar, J. Kumar and D. London, \textsl{arXiv:1704.07397 [hep-ph]}

\bibitem{3.5keV} \textsl{https://physics.aps.org/articles/v7/128 and also Astrophys. J. 789, 13 (2014) and  Phys. Rev. Lett. 113, 251301 (2014).}


\end{thebibliography}
\end{document}